\documentclass[a4paper,11pt]{article}
\pdfoutput=1 

\usepackage{jheppub} 

\usepackage[T1]{fontenc} 
\usepackage{amssymb}
\usepackage{amsmath}
\usepackage{amsthm}
\usepackage{epsfig}
\usepackage{graphicx}
\usepackage{subcaption}
\usepackage{bbold}
\usepackage{float}
\usepackage{multirow}
\usepackage{lineno}
\usepackage[normalem]{ulem} 
\usepackage{makeidx}
\usepackage{xspace}
\usepackage{wrapfig}
\usepackage{lastpage}
\usepackage{textcomp}
\usepackage{enumerate}
\usepackage{fancyhdr}
\usepackage{mathrsfs}
\usepackage[colorlinks=true,linkcolor=blue]{hyperref}
\usepackage{listings}
\usepackage{hyperref}
\usepackage{braket}
\usepackage{cancel}
\usepackage{feynmf}
\usepackage[compat=1.0.0]{tikz-feynman}
\usepackage{array}
\usepackage{tabularx}
\usepackage{ dsfont }
\numberwithin{equation}{section}
\usepackage{tikz-feynman}
\tikzfeynmanset{compat=1.0.0}
\usepackage{mathtools,slashed}
\usepackage[titletoc,title]{appendix}
\makeindex
\newcommand{\ga}{\alpha}
\newcommand{\gb}{\beta}
\newcommand{\gd}{\delta}

\newcommand{\gf}{\phi}
\newcommand{\gc}{\gamma}
\newcommand{\gh}{\eta}

\newcommand{\gl}{\lambda}
\newcommand{\gm}{\mu}
\newcommand{\gn}{\nu}
\newcommand{\go}{\theta}
\newcommand{\gp}{\pi}
\newcommand{\gq}{\psi}
\newcommand{\gr}{\rho}
\newcommand{\gs}{\sigma}
\newcommand{\gt}{\tau}

\newcommand{\gD}{\Delta}
\newcommand{\gF}{\Phi}
\newcommand{\gC}{\Gamma}

\newcommand{\gL}{\Lambda}

\newcommand{\gQ}{\Psi}

\newcommand{\gW}{\Omega}

\newcommand{\al}[3][2]{
\begin{align}
    #2 \label{#3}
\end{align}}

\newcommand{\pmatx}[1]{
\begin{pmatrix}
  #1
\end{pmatrix}}

\newcommand{\nal}[1]{
\begin{align*}
    #1
\end{align*}}
\newcommand{\prt}[1]{\left( #1\right)}
\newcommand{\crt}[1]{\left[ #1\right]}
\newcommand{\lrt}[1]{\left\{ #1\right\}}

\newcommand{\prtl}[1]{\left( #1\right.}
\newcommand{\crtl}[1]{\left[ #1\right.}
\newcommand{\lrtl}[1]{\left\{ #1\right.}

\newcommand{\prtr}[1]{\left. #1\right)}
\newcommand{\crtr}[1]{\left. #1\right]}
\newcommand{\lrtr}[1]{\left. #1\right\}}

\newcommand{\pdif}[3][2]{\dfrac{\partial #2}{\partial #3}}
\newcommand{\pddif}[3][2]{\dfrac{\partial^2 #2}{\partial #3^2}}
\newcommand{\ud}[3][2]{^{#3}_{#2}}

\newcommand{\half}{\frac{1}{2}}
\newcommand{\inv}[1]{\frac{1}{#1}}

\newcommand{\stxt}[1]{_{\textrm{#1}}}
\newcommand{\utxt}[1]{^{\textrm{#1}}}

\newcommand{\lag}{\mathcal{L}}

\newcommand{\dv}[1]{\partial_{#1}}
\newcommand{\idv}[1]{\partial^{#1}}

\newcommand{\vt}[3][2]{{#2}^{#3}}
\newcommand{\ft}[3][2]{{#2}_{#3}}

\newcommand{\nn}{\nonumber}

\newcommand{\hrm}[1]{{#1}^{\dagger}}

\newcommand{\e}[1]{e^{#1}}

\newcommand{\usm}[1]{\gs^{#1}}

\newcommand{\usmb}[1]{{\bar\gs}^{#1}}

\newcommand{\lsp}[3][2]{{#2}_{#3}}
\newcommand{\lspc}[3][2]{{#2}^{#3}}
\newcommand{\rsp}[3][2]{{\bar #2}^{\dot #3}}
\newcommand{\rspc}[3][2]{{\bar #2}_{\dot #3}}
\newcommand{\cev}[1]{\reflectbox{\ensuremath{\vec{\reflectbox{\ensuremath{#1}}}}}}

\title{\boldmath Functional renormalization group flows of $\mathcal{N}=1$ supersymmetric abelian gauge model with one chiral and one vector superfield}

\author[a]{Jeremy Echeverria,}
\author[b]{Maximiliano Binder,}
\author[c]{Iván Schmidt.}

\affiliation[a]{Instituto de Física, Pontificia Universidad Católica de Valparaíso, Casilla 4950, Valparaíso,
Chile}
\affiliation[b,c]{Departamento de Física, Universidad Técnica Federico Santa María,
y Centro Científico-Tecnológico de Valparaíso, Casilla 110-V, Valparaíso, Chile}
\emailAdd{jeremy.echeverria.p@mail.pucv.cl}
\emailAdd{maximiliano.binder@postgrado.usm.cl}
\emailAdd{ivan.schmidt@usm.cl}

\abstract{We apply the functional renormalization group approach to a $\mathcal{N}=1$ supersymmetric gauge model with one chiral superfield coupled to a vector $U(1)$ superfield. We find that the nonrenormalization theorem still works at leading order in the supercovariant derivative expansion of the fields. We also find the beta functions and we study the behavior of its fixed points in the local potential approximation. Regulators are also discussed.}

\begin{document} 
\maketitle
\flushbottom
\newpage

\section{Introduction}
\label{sec:1}
Supersymmetry is probably the most important formalism for achieving a unification of the fundamental interactions, in particular since it is the only symmetry known in which fermions and bosons are related by its transformations.  Although so far there has been no experimental confirmation of the Standard Model (SM) particles superpartners in nature, there still hope that indications will appear at higher energies, such as at the Large Hadron Collider. The present situation in particle  physics is that the Standard model (SM) has proven to work extremely well; nevertheless, it is not a unified theory and contains several unexplained parameters, so this makes us think that there must exist a most general theory that can explain these shortcomings \cite{drees1995implications}. Supersymmetry has been shown to be a possible solution, for example, to the hierarchy problem \cite{gildener1976gauge}, the smallness of the cosmological constant \cite{froggatt2006smallness} and the renormalizability of supergravity \cite{grisaru1976one}, among others, in addition to being a key ingredient of string theory.\\
Inside supersymmetric models, supersymmetric gauge theories have an special importance, being the main ingredient to build the minimal supersymmetric model (MSSM) \cite{bagger1996weak}, and being a powerful tool for approaching the quantum theory of gravity through the gauge-gravity duality \cite{maldacena1999large,witten1998anti,gubser1998gauge}. The non-perturbative analysis of supersymmetric gauge theories is important in order to investigate the mechanism of strong-coupling phenomena, some properties of condensation phenomena, aspects of the symmetry breaking as an approach to quantum gravity. Because of this it is clear that we need non-perturbative methods to study all aspects of explicit supersymmetry.\\
The functional renormalization group (FRG) has been an important non-perturbative tool to apply to a wide range of physical problems \cite{dupuis2021nonperturbative}. It consist in applying the functional approach of the Wilsonian renormalization group \cite{wilson1971renormalization,wilson1971renormalization2,wilson1974renormalization,fisher1998renormalization} and is based in an exact flow equation for a scale dependent effective average action (EAA) \cite{wetterich1991average,wetterich1993average,wetterich1993improvement,wetterich1993exact}. The FRG flow of supersymmetric models has been studied in supersymmetric quantum mechanics \cite{synatschke2009flow} and different types of Wess-Zumino models \cite{synatschke2010n,feldmann2016functional,feldmann2018critical}.\\
Abelian supersymmetric gauge theory is the simplest version of a supersymmetric gauge theory and it has been shown that these types of models can emerges from quantum phase transitions on topological insulators \cite{jian2017emergence}. In this work we study the beta functions and fixed points of an abelian $\mathcal{N}=1$ supersymmetric gauge theory with one chiral superfield in the FRG approach, using two types of regulators in the local potential approximation (LPA). The main goal is to provide a useful way to apply the FRG in supersymmetric gauge theories, being also the first application of it to supersymmetric theories including vector superfields coupled with chiral superfields. Also is the first step to the study of more elaborated models such as the MSSM.

\section{Chiral superfield coupled to $U(1)$ vector superfield}
\label{sec:2}
We start with the massless, non interacting Wess-Zumino model \cite{wess1974lagrangian} in $d=4$
\al{
\lag\stxt{WZ-free}=\dv{\gm}\hrm{\gf}\idv{\gm}\gf+i\bar\gq\bar\gs^\gm\dv{\gm}\gq+\hrm{F}F.
}{}
The most general interaction for this model has the form
\al{
\lag\stxt{int}=W'(\gf)F-\half W''(\gf)\gq\cdot\gq+\textrm{h.c},
}{}
where
\al{
W=\half m\gf^2+\inv{6}y\gf^3
}{}
is the superpotential, an holomorphic function of the scalar field. Here we only consider massive and yukawa-type terms (see chapter 5 of \cite{aitchison2007supersymmetry} for more details). The full lagrangian for the chiral part of our model is
\al{
\lag\stxt{chiral}=\dv{\gm}\hrm{\gf}\idv{\gm}\gf+i \bar\gq\bar\gs^\gm\dv{\gm}\gq+\hrm{F}F+\lrt{m\gf F+\half y\gf^2 F-\half m\gq\gq-\half y\gf\gq\gq+\textrm{h.c}}.
}{lag:chiral}
On the other hand, we need the gauge part of the lagrangian. We start with a massless abelian supersymmetric gauge theory with a gauge-fixing term
\al{
\lag\stxt{abelian-free}=-\inv{4}\ft{F}{\gm\gn}\vt{F}{\gm\gn}+i\bar\gl\bar\gs^\gm\dv{\gm}\gl+\half D^2+\inv{2\ga}\ft{A}{\gm}\vt{A}{\gm},
}{}
which we are going to couple it to (\ref{lag:chiral}) through the covariant derivative $D_\gm=\dv{\gm}+ig\ft{A}{\gm}$, where $g$ is the $U(1)$ coupling constant.  There is also necessary to add all possible renormalizable interactions between the fields that respect Lorentz, gauge and supersymmetric invariance. Following chapter 7.3 of \cite{aitchison2007supersymmetry}, the lagrangian of the model is
\al{
\lag=&D_{\gm}\hrm{\gf}D^{\gm}\gf+ i\bar\gq \bar\gs^\gm D_{\gm}\gq+\hrm{F}F-\inv{4}\ft{F}{\gm\gn}\vt{F}{\gm\gn}
+i \bar\gl \bar\gs^\gm\dv{\gm}\gl+\half D^2+\inv{2\ga}\ft{A}{\gm}\vt{A}{\gm}\nn\\
&+\lrt{m\gf F+\half y\gf^2 F-\half m\gq\gq-\half y\gf\gq\gq+\textrm{h.c}}+i\sqrt{2}g\crt{\hrm{\gf}\gq\gl-\gf\bar\gq\bar\gl}+g\hrm{\gf}\gf D.
}{eq:2.6}
It is useful to rewrite eq.(\ref{eq:2.6}) in terms of an "effective potential":
\al{
\lag=&\dv{\gm}\hrm{\gf}\idv{\gm}\gf+ i\bar\gq \bar\gs^\gm \dv{\gm}\gq+\hrm{F}F-\half\dv{\gm}\ft{A}{\gn}\idv{\gm}\vt{A}{\gn}+\half\dv{\gm}\ft{A}{\gm}\idv{\gn}\vt{A}{\gn}+i \bar\gl \bar\gs^\gm\dv{\gm}\gl+\half D^2+\inv{2\ga}\ft{A}{\gm}\vt{A}{\gm}\nn\\
&-\mathcal{U}\stxt{eff},
}{eq:effpot}
where
\al{
\mathcal{U}\stxt{eff}&=ig\dv{\gm}\hrm{\gf}\vt{A}{\gm}\gf-ig\ft{A}{\gm}\hrm{\gf}\idv{\gm}\gf-g^2\ft{A}{\gm}\vt{A}{\gm}\hrm{\gf}\gf+g \bar\gq \bar\gs^\gm\ft{A}{\gm}\gq\nn\\
&-\lrt{m\gf F+\half y\gf^2 F-\half m\gq\gq-\half y\gf\gq\gq+\textrm{h.c}}\nn\\
&-i\sqrt{2}g\crt{\hrm{\gf}\gq\gl-\gf\bar\gq\bar\gl}-g\hrm{\gf}\gf D.
}{}
The model can also be constructed from a lagrangian defined over the superspace (\cite{ferrara1974supergauge,wess1974supergauge}, chapter 4 of \cite{martin2010supersymmetry} and chapters 2 and 3 of \cite{bailin1994supersymmetric})
\al{
\lag=&\int d^2\go d^2\bar \go\gF^\dagger\e{2gV}\gF+\inv{32}\int d^2\go\vt{W}{\ga}\ft{W}{\ga}+\inv{\ga}\int d^2\go d^2\bar \go V^2+\int d^2\go \lrt{W(\gF)+\textrm{h.c}}
}{lagr}
where $\gF(x,\go,\hrm{\go})$ is defined as a chiral superfield and satisfies the conditions
\al{
\ft{\bar D}{\dot\ga}\gF=0\nn\\
\ft{D}{\ga}\hrm{\gF}=0
}{}
where
\al{
&\lsp{D}{\ga}=\pdif{}{\go^\ga}+i(\usm{\gm})_{\ga\dot\ga}\dv{\gm}\nn\\
&\rspc{D}{\ga}=-\pdif{}{\rsp{\go}{\ga}}-i\lspc{\go}{\ga}(\gs^\gm)_{\ga\dot\ga}\dv{\gm}.
}{}
The expansion in grassmann variables of the chiral superfield is
\al{
\gF(x,\go,\bar\go)=&\gf(x)+\sqrt{2}\go\gq(x)+\go\go F(x)+i\dv{\gm}\gf(x)\go\usm{\gm}\bar\go-\frac{i}{\sqrt{2}}\go\go\dv{\gm}\gq(x)\usm{\gm}\bar\go-\inv{4}\dv{\gm}\idv{\gm}\gf(x)\go\go\bar\go\bar\go.
}{}
Moreover, $W(\gF)$ is a superfunction of the chiral superfields that gives us the superpotential
\al{
W(\gF)=\half m\gF^2+\inv{6}y\gF^3.
}{}
On the other hand, $V(x,\go,\hrm{\go})$ is defined as a vector superfield and satisfies the condition
\al{
V=V^\dagger.
}{}
This superfield can be transformed through
\al{
V\to V+i\prt{\gW-\hrm{\gW}},
}{supergauge}
where $\gW$ is a chiral superfield. The transformation (\ref{supergauge}) is called a $U(1)$ supergauge transformation and we can semi-fix it such that it only holds for the usual gauge transformations. This is known as Wess-Zumino gauge \cite{ferrara1974supergauge} and its expansion in grassmann variables is
\al{
V\stxt{W-Z gauge}=\go\gs^\gm\bar\go\ft{A}{\gm}+i\go\go\bar\go\bar\gl-i\bar\go\bar\go\go\gl+\half \go\go\bar\go\bar\go D.
}{wzgauge}
Finally, $W_\ga$ is the gauge invariant abelian field strength superfield
\al{
\lsp{W}{\ga}\equiv \bar D^2\lsp{D}{\ga} V.
}{}
Using (\ref{wzgauge}), we can expand $\e{2gV}$ until $V^2$ terms (see section 4.8 of \cite{martin2010supersymmetry} for more details) and rewrite (\ref{lagr}) in a more useful form for our proposes
\al{
\lag=&\int d^2\go d^2\bar \go\gF^\dagger\gF+2g\int d^2\go d^2\bar \go\gF^\dagger V\gF+4g^2\int d^2\go d^2\bar \go\gF^\dagger V^2\gF\nn\\
&+\inv{32}\int d^2\go\vt{W}{\ga}\ft{W}{\ga}-\inv{\ga}\int d^2\go d^2\bar \go V^2+\int d^2\go \lrt{W(\gF)+\textrm{h.c}}\nn\\
=&\int d^2\go d^2\bar \go\gF^\dagger\gF+\inv{32}\int d^2\go\vt{W}{\ga}\ft{W}{\ga}-\inv{\ga}\int d^2\go d^2\bar \go V^2-U(\gF,V),
}{model}
with
\al{
U(\gF,V)&=-2g\int d^2\go d^2\bar \go\gF^\dagger V\gF-4g^2\int d^2\go d^2\bar \go\gF^\dagger V^2\gF-\int d^2\go \lrt{W(\gF)+\textrm{h.c}}
}{}
the effective potential.

\section{The functional renormalization group}
\label{sec:3}
The dependence on the scale of the model constants and operators will be threaded within the frame of functional renormalization group. This is constructed through the flow equation of the effective average action $\gC_k$, an action functional that is scale dependent and interpolates between the classical action at microscopic scales and the quantum effective action \cite{wetterich1991average}. On a certain scale of energy, where the system does not receive contributions of energy fluctuations, we introduce the energy cutoff scale $\gL$, and when the effective average action approximates $\gL$ we shall obtain the classical action, i.e the fluctuations may be frozen
\al{
\gC_{k\to \gL}=S.
}{5}
On the other hand, if the effective average action approximates to $0$, all fluctuations should be consider, i.e it will be
\al{
\gC_{k\to 0}=\gC.
}{}
The flow on the energy scale of the effective average action is obtained from the Wetterich equation \cite{wetterich1993exact}
\al{
\dv{t}\gC_k=\half\textrm{STr}\lrt{\crt{\gC\ud{k}{(2)}+R_k}^{-1}\dv{t}R_k},
}{wetterich eq}
where $t=\ln{\frac{k}{\gL}}$ and the supertrace denotes sum over all types of indices and momentum 
integration. $\gC\ud{k}{(2)}$ is the second functional derivative of the effective average action with respect to the fields\footnote{in momentum space.}:
\al{
\crt{\gC\ud{k}{(2)}}_{ij}(p,q)=\frac{\vec{\gd}}{\gd\gQ_i(p)}\gC_k\frac{\cev{\gd}}{\gd\gQ_j(q)}
}{}
where $\gQ$ is a collection of fields, and for this case, we use the arrangement
\al{
&\gQ^\dagger=(\hrm{\gf}(q),\gf(-q),\hrm{F}(q),F(-q),D(q),\ft{A}{\gm}(q),\bar\gq(q),\gq^T(-q),\bar\gl(q),\gl^T(-q))\nn\\
&\gQ=(\gf(q),\hrm{\gf}(-q),F(q),\hrm{F}(-q),D(-q),\ft{A}{\gn}(-q),\gq(q),\bar\gq^T(-q),\gl(q),\bar\gl^T(-q)).
}{}
On the other hand, $R_k=\gD S\ud{k}{(2)}$ is a matrix containing all regulator functions and it has the form
\al{
\gD S_k=\int d^4q \gQ^\dagger(q)R_k\gQ(q).
}{}

\section{Supersymmetric regulators}
\label{sec:4}
The goal of using the functional renormalization group is that it preserve supersymmetry in all steps of its implementation. To guarantee this we need to add a supersymmetric regulating term $\gD S_k$ in the formulation of EAA.\\
The types of regulators that can be used in supersymmetric theories in the local potential approximation(LPA) has been studied in, for example \cite{feldmann2016functional,feldmann2018critical}, and we are going to generalize them for vector superfields. Following \cite{feldmann2016functional} we use the regulator term
\al{
\gD S_k=&\int d^2\go d^2\bar \go\gF^\dagger\gr_2(D,\bar D)\gF+2g\int d^2\go d^2\bar \go\gF^\dagger \gr_G(D,\bar D)V\gF+4g^2\int d^2\go d^2\bar \go\gF^\dagger\gr_S(D,\bar D) V^2\gF\nn\\
&+\inv{32}\int d^2\go\vt{W}{\ga}\gt_2(D,\bar D)\ft{W}{\ga}+\lrt{\int d^4xd^2\go \gF\gr_1(D,\bar D)\gF+\textrm{h.c}}+\int d^4xd^2\go d^2 \bar\go V\gt_1(D,\bar D)V.
}{}
That can be simplified to the form
\al{
\gD S_k=&\int d^2\go d^2\bar \go\gF^\dagger r_2(\square)\gF+2g\int d^2\go d^2\bar \go\gF^\dagger r_G(\square)V\gF+4g^2\int d^2\go d^2\bar \go\gF^\dagger r_S(\square) V^2\gF\nn\\
&+\inv{32}\int d^2\go\vt{W}{\ga} t_2(\square)\ft{W}{\ga}+\lrt{\int d^4xd^2\go \gF r_1(\square)\gF+\textrm{h.c}}+\int d^4xd^2\go d^2 \bar\go V t_1(\square)V.
}{reg}
The reason for using distinct names for the regulator functions in each term is due to the dimensionless process, as each function is accompanied by different types of fields which will contain different wave functions renormalization. The explicit form of the $R_k$ operator is shown in Appendix \ref{sec:B}.\\
According to \cite{feldmann2016functional}, we use two types of regulators:
\begin{itemize}
    \item Callan Symanzik regulator(CS): $r_1=t_1=1$ and $r_2=t_2=r_G=r_S=0$.
    \item Litim-type regulator(LT): $r_1=t_1=0$ and $r_2=t_2=r_G=r_S=\prt{\dfrac{1}{q}-1}\go(1-q^2)$
\end{itemize}
for dimensionless and renormalized regulators.

\section{Supersymmetric and gauge invariant LPA' truncation}
\label{sec:5}
We study the FRG flows of the LPA' truncation for this model, that is, consider the effective potential (\ref{eq:effpot}) as scale dependent. The truncation used for $\gC_k$ is
\al{
\gC_k=&\int d^4x\lrtl{Z_{\gF}\dv{\gm}\hrm{\gf}\idv{\gm}\gf+ iZ_{\gF}\bar\gq \bar\gs^\gm \dv{\gm}\gq+Z_{\gF}\hrm{F}F}-iZ_{G}g_k\dv{\gm}\hrm{\gf}\vt{A}{\gm}\gf+iZ_{G}g_k\ft{A}{\gm}\hrm{\gf}\idv{\gm}\gf\nn\\
&+Z_{S}g_k^2\ft{A}{\gm}\vt{A}{\gm}\hrm{\gf}\gf-Z_{G}g_k \bar\gq \bar\gs^\gm\ft{A}{\gm}\gq+i\sqrt{2}Z_{G}g_k\crt{\hrm{\gf}\gq\gl-\gf\bar\gq\bar\gl}+Z_{G}g_k\hrm{\gf}\gf D\nn\\
&+Z_{V}\prtl{-\half\dv{\gm}\ft{A}{\gn}\idv{\gm}\vt{A}{\gn}+\half\dv{\gm}\vt{A}{\gm}\dv{\gn}\vt{A}{\gn}}+\inv{2\ga}\ft{A}{\gm}\vt{A}{\gm}\prtr{+i \bar\gl \bar\gs^\gm\dv{\gm}\gl+\half D^2}\nn\\
&+\lrtr{\crt{\pdif{W_k}{\gf}F-\half\pddif{W_k}{\gf}\gq\gq+\textrm{h.c}}},
}{act}
or in superspace formulation
\al{
\gC_k=&Z_\gF\int d^2\go d^2\bar \go\gF^\dagger\gF+\frac{Z_V}{32}\int d^2\go\vt{W}{\ga}\ft{W}{\ga}-\frac{Z_V}{\ga}\int d^2\go d^2\bar \go V^2-U_k(\gF,V)\nn\\
=&Z_\gF\int d^2\go d^2\bar \go\gF^\dagger\gF+2g_kZ_G\int d^2\go d^2\bar \go\gF^\dagger V\gF+4g_k^2Z_S\int d^2\go d^2\bar \go\gF^\dagger V^2\gF\nn\\
&+\frac{Z_V}{32}\int d^2\go\vt{W}{\ga}\ft{W}{\ga}-\frac{Z_V}{\ga}\int d^2\go d^2\bar \go V^2+\int d^2\go \lrt{W_k(\gF)+\textrm{h.c}}.
}{}
where ($Z_{\gF}$, $Z_{V}$, $Z_{g}$) are the wave functions renormalization and does not depend on the energy scale. Besides we have
\al{
W_k=&\half m_k\gf^2+\inv{6}y_k\gf^3,\nn\\
Z_{G}=&Z_{\gF}Z_{g}\sqrt{Z_{V}} \quad\textrm{and}\nn\\
Z_{S}=&Z_{\gF}Z^2_{g}Z_{V}.
}{c}
The explicit form of the operator $\gC^{(2)}$ is shown in Appendix \ref{sec:B}.

\section{Flow equations}
\label{sec:6}
In this section we present the results for the non-renormalization theorem and the beta functions for the dimensionless gauge ($g_k$) and Yukawa ($y_k$) couplings, in addition to the dimensionless mass ($m_k$), in the LPA' truncation. First, we write (\ref{act}) and (\ref{reg}) in momentum space; and then, because the wave functions renormalization do not depend on the energy scale, we can restrict all fields to real constants $\gQ_i(q)=(\sqrt{2\gp})^4\gQ_i\gd(q)$. Finally, in Appendix \ref{sec:A} we show the dimensionless and renormalized form of the functions, which are used to write the beta functions in a dimensionless form.
\subsection{Non-renormalization theorem}
If we replace (\ref{act}) in the left side of (\ref{wetterich eq}) and we set $\gq=\bar\gq=\gl=\bar\gl=D=A=0$ we obtain
\al{
\dv{t}\gC_k=(\sqrt{2\gp})^4\gd(0)\dv{t}\crt{Z_{\gF}\hrm{F}F+\pdif{W_k}{\gf}F+\pdif{\hrm{W}_k}{\hrm{\gf}}\hrm{F}},
}{a}
thus, we can obtain the flow of the superpotential taking the $F$ or $\hrm{F}$ derivative and setting $F=\hrm{F}=0$ respectively. Doing this we obtain
\al{
\dv{t}W=0,
}{}
which confirms that the non-renormalization theorem works in the LPA' approximation.
\subsection{Beta Functions and Fixed Points}
In $d$ dimensions the beta functions are
\al{
\gb_{g_k}&=\prt{\gh_{g}+\frac{d-4}{2}}g_k\nn\\
\gb_{y_k}&=\prt{-\frac{3}{2}\gh_{\gF}-\frac{d-4}{2}}y_k,\nn\\
\gb_{m_k}&=\prt{\gh_{\gF}-1}m_k,
}{}
where $\gh_\mathcal{O}=-\dv{t}\ln{Z_\mathcal{O}}$ is the LPA' anomalous dimension of the operator $\mathcal{O}$ and $\gb_\mathcal{O}=\dv{t}\mathcal{O}$ is the beta function. We compute the anomalous dimensions in the same way as before but setting $\gq=\hrm{\gq}=F=\hrm{F}=A=\gl=\hrm{\gl}=0$ in (\ref{wetterich eq})
\al{
\dv{t}\gC_k=(\sqrt{2\gp})^4\gd(0)\dv{t}\crt{g_kZ_{G}\hrm{\gf}\gf D+\frac{Z_{V}}{2}D^2}.
}{b}
We compute the anomalous dimensions using
\begin{itemize}
    \item $\gh_{\gF}$ taking the $\gf,\hrm{\gf}$ derivatives of (\ref{a}) at $\gf=\hrm{\gf}=0$,
    \item $\gh_{V}$ taking the $D$ second derivative of (\ref{b}) at $D=0$,
    \item $\gh_{G}$ taking the $\gf,\hrm{\gf},D$ derivatives of (\ref{b}) at $\gf=\hrm{\gf}=D=0$
\end{itemize}
and using (\ref{c}) to write
\al{
\gh_{g}=\gh_{G}-\gh_{\gF}-\half\gh_{V}.
}{}

\subsubsection{Callan-Symanzik regulator}
Using the CS regulator the beta functions in $d=4$ are
\al{
\gb\utxt{CS}_{y_{k}}&=\frac{24\sqrt{2}\gp^2y_{k}^3}{96+192\gp+48\gp^3+6\gp^4+3(2+\gp)^4m_{k}-16\gp^2(\sqrt{2}y_{k}^2-9)},\nn\\
\gb\utxt{CS}_{m_{k}}&=-\frac{3(2+\gp)^4m_{k}(m_{k}+2)}{96+192\gp+48\gp^3+6\gp^4+3(2+\gp)^4m_{k}-16\gp^2(\sqrt{2}y_{k}^2-9)},\nn\\
\gb\utxt{CS}_{g_{k}}&=\frac{48\sqrt{2}\gp^2g_{k}(g_{k}^2+y_{k}^2)}{48+96\gp+24\gp^3+3\gp^4+3(2+\gp)^4m_{k}-8\gp^2(\sqrt{2}y^2-9)}.
}{}
The explicit form of the anomalous dimensions in $d$ dimensions are shown in Appendix \ref{apx:C}.\\
In this case we find that the system only exhibit gaussian fixed points ($g_k=y_k=m_k=0$).

\subsubsection{Litim-Type regulator}
\begin{figure}[h]
\begin{center}
\begin{subfigure}{0.48\textwidth}
\includegraphics[width=\textwidth]{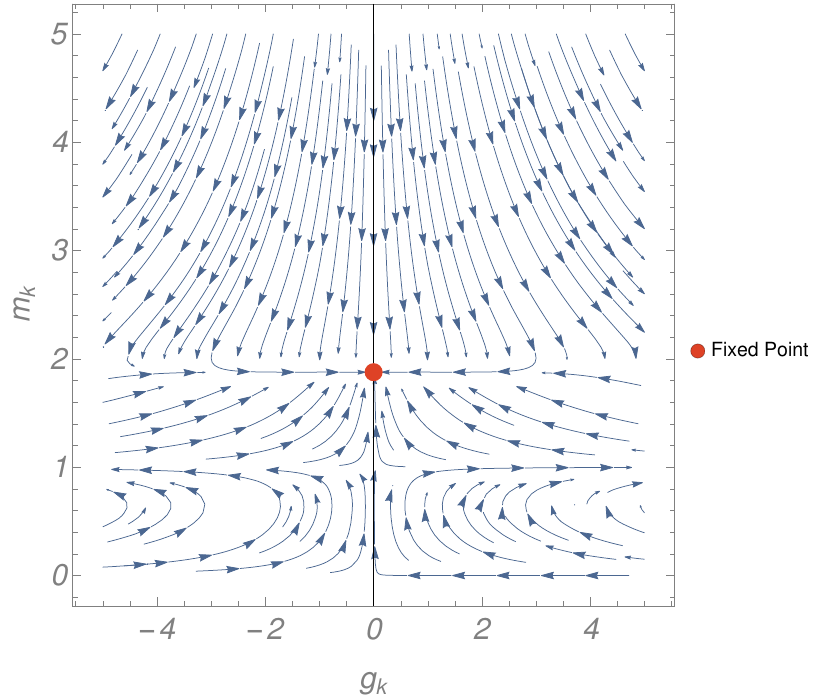}
\caption{$y_k=10$.}
\label{fig:y10}
\end{subfigure}
\begin{subfigure}{0.48\textwidth}
\includegraphics[width=\textwidth]{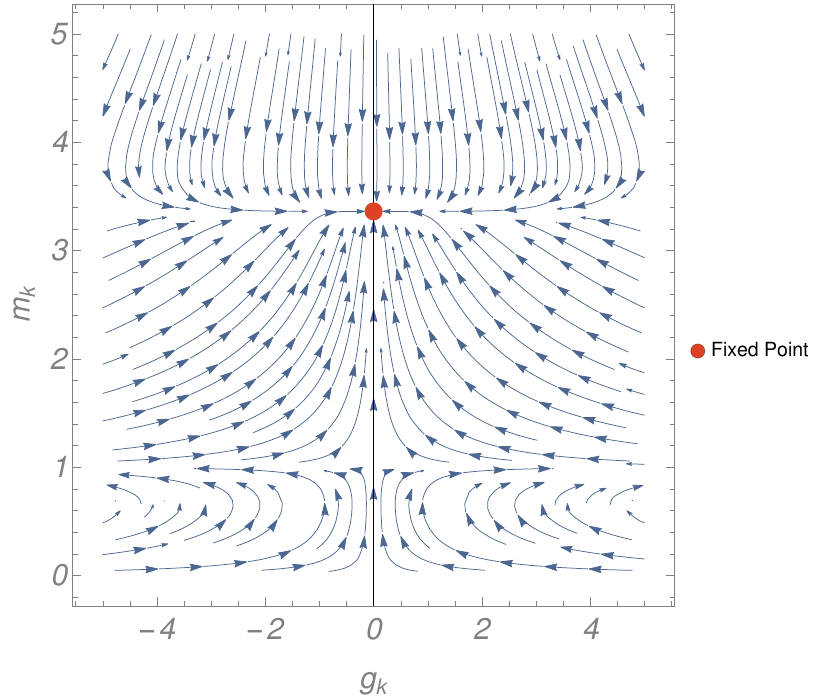}
\caption{$y_k=50$.}
\label{fig:y50}
\end{subfigure}
\end{center}
\caption{Flow Diagrams in the $g_k-m_k$ projection for different values of $y_k$. Here $m(g)$ is represented in the $y(x)$ axis.}
\label{fig:flow}
\end{figure}

\begin{figure}[h]
\begin{center}
\includegraphics[width=0.5\textwidth]{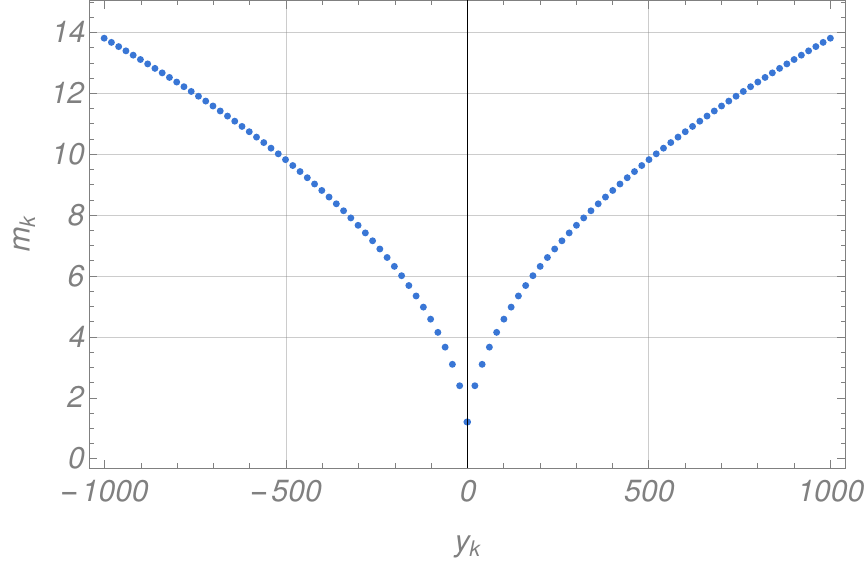}
\end{center}
\caption{Fixed point, for different values of $m_k$ and $y_k$.}
\label{fig:fixed}
\end{figure}

Using the LT regulator the beta functions in $d=4$ are
\al{
\gb\utxt{LT}_{y_{k}}&=-\frac{36\sqrt{2}\gp^\frac{5}{2}(m_k^2+1)y_k^3}{9\sqrt{\gp}(m_k^2-1)^3+8\sqrt{2}\gp^\frac{5}{2}(m_k^2+1)y_k^2}\nn\\
\gb\utxt{LT}_{m_{k}}&=\frac{\sqrt{\gp}m_k\prt{16\sqrt{2}\gp^2y_k^2(m_k^2+1)-9(2+\gp)^4(m_k^2-1)^3}}{9\sqrt{\gp}(m_k^2-1)^3+8\sqrt{2}\gp^\frac{5}{2}(m_k^2+1)y_k^2}\nn\\
\gb\utxt{LT}_{g_{k}}&=\frac{g_k\prt{m_k^2(m_k^2-1)^3\prt{144+\gp+72\gp^3+9\gp^4+216\gp^2}-4\gp^2\sqrt{2}(7m_k^4+4m_k^2-3)}}{(m_k^2-1)^4\prt{144+\gp+72\gp^3+9\gp^4+216\gp^2}+8\sqrt{2}\gp^2(m_k^4-1)y_k^2}.
}{}
The explicit form of the anomalous dimensions are shown in Appendix \ref{apx:C}. In this case we find non-gaussian fixed points for $g_k=0$, $y_k\neq 0$ and $m_k\neq 0$. We can see in fig.\ref{fig:flow} that there exists an unstable fixed point at $m_k=1$ and a stable one, which changes continuously its position when $y_k$ grows. We can see in fig.\ref{fig:fixed} the continuous change in the fixed point value for different values of $y_k$.
\newpage

\section{Comments and conclusions}
\label{sec:7}
We have obtained the beta functions for an abelian supersymmetric gauge theory in LPA' with a supersymmetric regularization for two different regulator functions. We found that the non-renormalization theorem still works when we couple the chiral superfield with a vector superfield in LPA'. For the case of Callan-Symanzik-type (CS) regulator we found that there are no non-gaussian fixed points, i.e, it does not exist a point $\lrt{g_*,y_*,m_*}\neq\lrt{0,0,0}$ for wich $\gb_g=\gb_y=\gb_m=0$. On the other hand, for the Litim-type (LT) regulator we found that there exists stable non-gaussian fixed points of the form $\lrt{g_*,m_*}=\lrt{0,m}$ and unestable non-gaussian fixed points $\lrt{g_*,m_*}=\lrt{0,1}$ for any value of $y_*$. In the case $y=0$ only exists the unestable fixed point.\\
Clearly both types of regularizations give completely different results.  In the study of the Wess-Zumino models the CS regulator has been shown to be a non-optimized regulator \cite{gies2009supersymmetry} in comparison with the optimized LT regulator \cite{litim2001optimized}. In fact, a mass-type regulator for the vector superfield in the Wess-Zumino gauge as CS regulator seems not to be a good option because it does not work to regularized $\gl$ and $D$ fields, because the Wess-Zumino gauge allows that quadratic terms in vector superfield to be proportional to $A^2$.\\
The most important contribution of this work has been to present a useful way to study beta functions of supersymmetric gauge theories in the non-perturbative regime, expanding previous works in Wess-Zumino models to include vector superfields, using convenient regulator functions. We have studied an abelian gauge theory which involves only one chiral superfield in a LPA',  focusing in the application of the FRG in this type of models, which in future work can be generalized to models with more field content and next order truncations in a superderivative expansion.


\acknowledgments

This research was partially supported by project Proyecto ANID PIA/APOYO AFB180002 (Chile). J.E has been funded by Beca doctorado nacional Chile ANID N°21212381 and project FONDECYT N°1180232.

\begin{appendices}


\section{Conventions and dimensionless-renormalized quantities}
\label{sec:A}
For momentum space we use the convention
\al{
F(q)=\dfrac{1}{\prt{\sqrt{2\gp}}^d}\displaystyle\int d^dx F(x)\e{i\ft{q}{\gm}\vt{x}{\gm}}.
}{}
In order to write anomalous dimensions and beta functions in term of dimensionless and renormalized quantities we defined
\al{
m_k=&Z_\gF km_{k,R}\nn\\
y_k=&\sqrt{(Z_\gF)^3}k^{\frac{d-4}{2}}y_{k,R}\nn\\
g_k=&Z_gk^{\frac{d-4}{2}}g_{k,R},
}{}
also we need the regulator functions in a dimensionless form, for that we redefine
\al{
r_1=&Z_\gF kr_{1,R}\nn\\
t_1=&Z_V kt_{1,R}\nn\\
r_2=&Z_\gF r_{2,R}\nn\\
t_2=&Z_V t_{2,R}\nn\\
r_G=&Z_gZ_\gF\sqrt{Z_V} r_{G,R}\nn\\
r_S=&Z_\gF Z_V Z_g^2 r_{S,R}.
}{}
It is convenient to solve the integrals in Appendix \ref{sec:C} change the integration variable by $q_R\to\dfrac{q}{k}$ and defining $r'_{i,R}(q_R^2)=r_{i,R}(k^2q_R^2)$. Finally, we omit all $R$ subscripts and primes.

\section{Operators LPA'}
\label{sec:B}
The way to found the explicit form of the operators $R_k$ and $\gC\ud{k}{(2)}$ is, in first place, write the regulator \ref{reg} and the action \ref{act} in momentum space, then restrict all fields to real constants $\gQ_i(q)=(\sqrt{2\gp})^4\gQ_i\gd(q)$ and compute the matrix $R_k$ and the second derivatives of EAA.\\
In our case, the explicit form of operators are
\al{
R_k=\pmatx{R_{kBB} & R_{kBF}\\
R_{kFB} & R_{k FF}},
}{}
where
\nal{
R_{kBB}=&\pmatx{r_2q^2+g_{k}r_GD+g_{k}r_SA^2 & 0 & 0 & 2r_1 & g_{k}r_G\gf & 2g_{k}^2r_S\gf A\\
               0 & r_2q^2+g_{k}r_GD+g_{k}r_SA^2 & 2r_1 & 0 & g_{k}r_G\hrm{\gf} & 2g_{k}^2r_S\hrm{\gf}A \\
         0 & 2r_1 & r_2 & 0 & 0 & 0\\
         2r_1 & 0 & 0 & r_2 & 0 & 0\\
         g_{k}r_G\hrm{\gf} & g_{k}r_G\gf & 0 & 0 & t_2 & 0\\
         2g_{k}^2r_S\hrm{\gf}A & 2g_{k}^2r_S\gf A & 0 & 0 & 0 & t_1+\frac{q^2}{\ga}t_2+2g_k^2r_S \gf\hrm{\gf}
         }\nn\\
R_{kFB}=&\pmatx{
            0 & 0 & \sqrt{2}g_kr_G\gq^T\gs^2 & 0\\
            0 & \sqrt{2}g_k r_G \bar\gl\gs^2 & 0 & 0\\
            0 & 0 & 0 & 0\\
            0 & 0 & 0 & 0\\
            0 & 0 & 0 & 0\\
            g_kr_G\bar\gs^\gm\bar\gq & g_kr_G\bar\gs^\gm\gq^T & 0 & 0
            }\nn\\
R_{kBF}=&\pmatx{
            0 & 0 & 0 & 0 & 0 & g_kr_G\bar\gs^\gm\gq\\
            0 & \sqrt{2}g_k r_G \gl\gs^2 & 0 & 0 & 0 & g_kr_G\bar\gs^\gm\bar\gq^T\\
            \sqrt{2}g_k r_G \bar\gq^T\gs^2 & 0 & 0 & 0 & 0 & 0\\
            0 & 0 & 0 & 0 & 0 & 0
            }\nn\\
R_{kFF}=&\pmatx{ r_2\ft{q}{\gm}\usmb{\gm}+g_kr_G\ft{A}{\gm}\usmb{\gm} & ir_1\gs^2 & 0 & 0 \\
               -ir_1\gs^2 & r_2\ft{q}{\gm}\usmb{T\gm}+g_kr_G\ft{A}{\gm}\usmb{T\gm} & \sqrt{2}g_{k}r_G\hrm{\gf}\gs^2 & 0\\
               0 & \sqrt{2}g_kr_G\gf\gs^2 & t_2\ft{q}{\gm}\usmb{\gm } & 0\\
               0 & 0 & 0 & t_2\ft{q}{\gm}\usmb{\gm T}}.
}
In the same way
\al{
\gC_k=\pmatx{\gC_{kBB} & \gC_{kBF}\\
\gC_{kFB} & \gC_{k FF}},
}{}
with
\al{
\gC_{k BB}=&\pmatx{\gc & y_k\hrm{F} & 0 & m_k+y_k\hrm{\gf} & g_{k}Z_G\gf & 2g_{k}^2Z_S\gf A\\
y_kF & \gc & m_k+y_k\gf & 0 & g_{k}Z_G\hrm{\gf} & 2g_{k}^2Z_S\hrm{\gf}A\\
0 & m_k+y_k\hrm{\gf} & Z_{\gF} & 0 & 0 & 0\\
m_k+y_k\gf & 0 & 0 & Z_{\gF} & 0 & 0\\
g_{k}Z_G\hrm{\gf} & g_{k}Z_G\gf & 0 & 0 & Z_{V} & 0\\
2g_{k}^2Z_S\hrm{\gf}A & 2g_{k}^2Z_S\gf A & 0 & 0 & 0 & \frac{q^2}{\ga}Z_V+2g_k^2Z_S\gf\hrm{\gf}}\nn\\
\gC_{kFB}=&\pmatx{
            0 & -\frac{iy_k\bar\gq}{8\gp^2}\gs^2 & \sqrt{2}g_kZ_G\gq^T\gs^2 & 0\\
            \frac{iy_k\gq^T}{8\gp^2}\gs^2 & \sqrt{2}g_kZ_G\bar\gl\gs^2 & 0 & 0\\
            0 & 0 & 0 & 0\\
            0 & 0 & 0 & 0\\
            0 & 0 & 0 & 0\\
            g_kZ_G\bar\gq\bar\gs^\gm & g_kZ_G \gq^T\bar\gs^\gm & 0 & 0
            }\nn\\
\gC_{k BF}=&\pmatx{
            0 & -\frac{iy_k\bar\gq^T}{8\gp^2}\gs^2 & 0 & 0 & 0 & g_kZ_G\gq\bar\gs^\gm\\
            \frac{iy_k\gq}{8\gp^2}\gs^2 & \sqrt{2}g_kZ_G\gl\gs^2 & 0 & 0 & 0 & g_kZ_G\bar\gq^T\bar\gs^\gm\\
            \sqrt{2}g_kZ_G\bar\gq^T\gs^2 & 0 & 0 & 0 & 0 & 0\\
            0 & 0 & 0 & 0 & 0 & 0
            }\nn\\
\gC_{kFF}=&\pmatx{Z_\gf\ft{q}{\gm}\usmb{\gm}+g_kZ_G\ft{A}{\gm}\usmb{\gm}  & -\dfrac{i(m_k+y_k\hrm{\gf})}{4\gp^2}\gs^2 & 0 & 0\\
\dfrac{i(m_k+y_k\gf)}{4\gp^2}\gs^2 & Z_\gf\ft{q}{\gm}\usmb{T\gm}+g_kZ_G\ft{A}{\gm}\usmb{T\gm} & \sqrt{2}g_{k}Z_{G}\hrm{\gf}\gs^2 & 0\\
0 & \sqrt{2}g_{k}Z_{G}\gf\gs^2 & Z_{V}q_\gm\usmb{\gm} & 0\\
0 & 0 & 0 & Z_{V}q_\gm\usmb{\gm T} }.
}{}
Where
\al{
\gc=&Z_{\gF}q^2+Z_{S}g_{k}^2A^2
+Z_{G}g_kD.
}{}

\section{Anomalous Dimensions LPA'}
\label{apx:C}
In $d$ dimensions, the explicit form of the dimensionless anomalous dimensions in both regulators are
\al{
\gh_{\gF}^\textrm{CS}=&-\int\ud{0}{\infty} dq\frac{\gW_d}{(2\gp)^d}q^{d-1}\frac{4y_k^2\ga_1(1-\gh^\textrm{CS}_{\gF}+\dv{t}-q\dv{q})r_1}{(\ga_1^2-q^2)^3}\nn\\
\gh_{V}^\textrm{CS}=&-\int\ud{0}{\infty} dq\frac{\gW_d}{(2\gp)^d}q^{d-1}\frac{8g_k^2\ga_1(1-\gh^\textrm{CS}_{\gF}+\dv{t}-q\dv{q})r_1}{(\ga_1^2-q^2)^3}\nn\\
\gh_{g}^\textrm{CS}=&\int\ud{0}{\infty} dq\frac{\gW_d}{(2\gp)^d}q^{d-1}\frac{4\ga_1\crt{3g_k^2\prt{\ga_1^2-q^2}-y_k^2\prt{\ga_1^2+5q^2}}(1-\gh^\textrm{CS}_{\gF}+\dv{t}-q\dv{q})r_1}{(\ga_1^2-q^2)^4},
}{}
where
\nal{
\ga_1=m+2r_1.
}
\al{
\gh_{\gF}^\textrm{LT}=&-\int\ud{0}{1} dq\frac{\gW_d}{(2\gp)^d}q^{d-1}\frac{y_k^2\ga_2\prt{\ga_2^2q^2+m_k^2}(-\gh_{\gF}^\textrm{LT}+\dv{t}-q\dv{q})r_2}{(\ga_2^2q^2-m_k^2)^3}\nn\\
\gh_{V}^\textrm{LT}=&-\int\ud{0}{1} dq\frac{\gW_d}{(2\gp)^d}q^{d-1}\inv{(\ga_2^2q^2-m_k^2)^3}\lrtl{2g_k^2\ga_2\ga_G\crtl{\ga_G\prt{\ga_2^2q^2+m_k^2}(-\gh_{\gF}^\textrm{LT}+\dv{t}-q\dv{q})r_2}}\nn\\
&\lrtr{\crtr{-\ga_2\prt{\ga_2^2q^2-m_k^2}(-\gh_{g}^\textrm{LT}-\gh_{\gF}^\textrm{LT}-\tfrac{1}{2}\gh_{V}^\textrm{LT}+\dv{t}-q\dv{q})r_G}}\nn\\
\gh_{g}^\textrm{LT}=&-\int\ud{0}{1} dq\frac{\gW_d}{(2\gp)^d}q^{d-1}\inv{(\ga_2^2q^2-m_k^2)^4\ga_t^2}\lrtl{\ga_t\crtl{y_k^2\ga_t\prt{m_k^4\prt{r_2-r_G}-8m_k^2q^2\ga_2^2\ga_G-q^4\ga_2^4\prt{4+r_2+3r_G}}}}\nn\\
&-\crtr{g_k^2\ga_2\prt{\ga_2^4q^4-m_k^4}\ga_G^2\prt{3+2r_G+t_2}}(-\gh_{\gF}^\textrm{LT}+\dv{t}-q\dv{q})r_2\nn\\
&-\ga_2\prt{\ga_2^2-m_k^2}\crtl{g_k^2\ga_2\prt{\ga_2^2q^2-m_k^2}\ga_G^3(-\gh_{V}^\textrm{LT}+\dv{t}-q\dv{q})t_2}\nn\\
&-\ga_t\prt{g_k^2\ga_2\prt{\ga_2^2q^2-m_k^2}\ga_G\prt{4+3r_G+t_2}-y_k^2\ga_t\prt{\ga_2^2q^2+m_k^2}}\times\nn\\
&\times\lrtr{\crtr{(-\gh_{g}^\textrm{LT}-\gh_{\gF}^\textrm{LT}-\tfrac{1}{2}\gh_{V}^\textrm{LT}+\dv{t}-q\dv{q})r_G}},
}{}
where $\ga_2=r_2+1$, $\ga_G=r_G+1$, $\ga_t=t_2+1$ and $\gW_d=\dfrac{(2\gp)^{\frac{d+1}{2}}}{\gC\crt{\frac{d+1}{2}}}$ is the surface of a $d-$dimensional sphere.

\end{appendices}


\bibliographystyle{unsrt}
\bibliography{refs}

\providecommand{\noopsort}[1]{}\providecommand{\singleletter}[1]{#1}%
\begin{thebibliography}{10}

\bibitem{drees1995implications}
Manuel Drees and Stephen~P Martin.
\newblock {I}mplications of {SUSY} model building.
\newblock {\em arXiv preprint hep-ph/9504324}, 1995.

\bibitem{gildener1976gauge}
Eldad Gildener.
\newblock {G}auge-symmetry hierarchies.
\newblock {\em Physical Review D}, 14(6):1667, 1976.

\bibitem{froggatt2006smallness}
C~Froggatt, R~Nevzorov, and HB~Nielsen.
\newblock {O}n the smallness of the cosmological constant in {SUGRA} models.
\newblock {\em Nuclear Physics B}, 743(1-2):133--152, 2006.

\bibitem{grisaru1976one}
Marcus~T Grisaru, P~Van~Nieuwenhuizen, and JAM Vermaseren.
\newblock {O}ne-loop renormalizability of pure supergravity and of
  {M}axwell-{E}instein theory in extended supergravity.
\newblock {\em Physical Review Letters}, 37(25):1662, 1976.

\bibitem{bagger1996weak}
Jonathan~A Bagger.
\newblock {W}eak-scale supersymmetry: theory and practice.
\newblock {\em arXiv preprint hep-ph/9604232}, 1996.

\bibitem{maldacena1999large}
Juan Maldacena.
\newblock {T}he large-{N} limit of superconformal field theories and
  supergravity.
\newblock {\em International journal of theoretical physics}, 38(4):1113--1133,
  1999.

\bibitem{witten1998anti}
Edward Witten.
\newblock {A}nti de {S}itter space and holography.
\newblock {\em arXiv preprint hep-th/9802150}, 1998.

\bibitem{gubser1998gauge}
Steven~S Gubser, Igor~R Klebanov, and Alexander~M Polyakov.
\newblock {G}auge theory correlators from non-critical string theory.
\newblock {\em Physics Letters B}, 428(1-2):105--114, 1998.

\bibitem{dupuis2021nonperturbative}
Nicolas Dupuis, L~Canet, Astrid Eichhorn, W~Metzner, Jan~M Pawlowski,
  M~Tissier, and N~Wschebor.
\newblock {T}he nonperturbative functional renormalization group and its
  applications.
\newblock {\em Physics Reports}, 910:1--114, 2021.

\bibitem{wilson1971renormalization}
Kenneth~G Wilson.
\newblock {R}enormalization group and critical phenomena. {I}.
  {R}enormalization group and the {K}adanoff scaling picture.
\newblock {\em Physical review B}, 4(9):3174, 1971.

\bibitem{wilson1971renormalization2}
Kenneth~G Wilson.
\newblock {R}enormalization group and critical phenomena. {II}. {P}hase-space
  cell analysis of critical behavior.
\newblock {\em Physical Review B}, 4(9):3184, 1971.

\bibitem{wilson1974renormalization}
Kenneth~G Wilson and John Kogut.
\newblock {T}he renormalization group and the $\gee$ expansion.
\newblock {\em Physics reports}, 12(2):75--199, 1974.

\bibitem{fisher1998renormalization}
Michael~E Fisher.
\newblock {R}enormalization group theory: {I}ts basis and formulation in
  statistical physics.
\newblock {\em Reviews of Modern Physics}, 70(2):653, 1998.

\bibitem{wetterich1991average}
Christof Wetterich.
\newblock {A}verage action and the renormalization group equations.
\newblock {\em Nuclear Physics B}, 352(3):529--584, 1991.

\bibitem{wetterich1993average}
Christof Wetterich.
\newblock {T}he average action for scalar fields near phase transitions.
\newblock {\em Zeitschrift f{\"u}r Physik C Particles and Fields},
  57(3):451--469, 1993.

\bibitem{wetterich1993improvement}
Ch~Wetterich.
\newblock {I}mprovement of the average action.
\newblock {\em Zeitschrift f{\"u}r Physik C Particles and Fields},
  60(3):461--469, 1993.

\bibitem{wetterich1993exact}
Christof Wetterich.
\newblock {E}xact evolution equation for the effective potential.
\newblock {\em Physics Letters B}, 301(1):90--94, 1993.

\bibitem{synatschke2009flow}
Franziska Synatschke, Georg Bergner, Holger Gies, and Andreas Wipf.
\newblock {F}low equation for supersymmetric quantum mechanics.
\newblock {\em Journal of High Energy Physics}, 2009(03):028, 2009.

\bibitem{synatschke2010n}
Franziska Synatschke, Jens Braun, and Andreas Wipf.
\newblock N= 1 {W}ess-{Z}umino model in d= 3 at zero and finite temperature.
\newblock {\em Physical Review D}, 81(12):125001, 2010.

\bibitem{feldmann2016functional}
Polina Feldmann.
\newblock {\em {F}unctional {R}enormalization {G}roup {A}pproach to the
  3-{D}imensional {N}= 2 {W}ess-{Z}umino {M}odel}.
\newblock PhD thesis, Faculty of Physics and Astronomy, Friedrich Schiller
  University Jena, 2016.

\bibitem{feldmann2018critical}
Polina Feldmann, Andreas Wipf, and Luca Zambelli.
\newblock Critical {W}ess-{Z}umino models with four supercharges in the
  functional renormalization group approach.
\newblock {\em Physical Review D}, 98(9):096005, 2018.

\bibitem{jian2017emergence}
Shao-Kai Jian, Chien-Hung Lin, Joseph Maciejko, and Hong Yao.
\newblock {E}mergence of supersymmetric quantum electrodynamics.
\newblock {\em Physical review letters}, 118(16):166802, 2017.

\bibitem{wess1974lagrangian}
Julius Wess and Bruno Zumino.
\newblock {A} {L}agrangian model invariant under supergauge transformations.
\newblock {\em Physics Letters B}, 49(1):52--54, 1974.

\bibitem{aitchison2007supersymmetry}
Ian Aitchison.
\newblock {\em {S}upersymmetry in particle physics: an elementary
  introduction}.
\newblock Cambridge University Press, 2007.

\bibitem{ferrara1974supergauge}
Sergio Ferrara and B~Zumino.
\newblock {S}upergauge invariant {Y}ang-{M}ills theories.
\newblock {\em Nuclear Physics B}, 79(3):413--421, 1974.

\bibitem{wess1974supergauge}
Julius Wess and Bruno Zumino.
\newblock {S}upergauge transformations in four dimensions.
\newblock {\em Nuclear Physics B}, 70(1):39--50, 1974.

\bibitem{martin2010supersymmetry}
Stephen~P Martin.
\newblock {A} supersymmetry primer.
\newblock In {\em Perspectives on supersymmetry II}, pages 1--153. World
  Scientific, 2010.

\bibitem{bailin1994supersymmetric}
David Bailin and Alexander Love.
\newblock {\em {S}upersymmetric gauge field theory and string theory}.
\newblock Taylor \& Francis, 1994.

\bibitem{gies2009supersymmetry}
Holger Gies, Franziska Synatschke, and Andreas Wipf.
\newblock {S}upersymmetry breaking as a quantum phase transition.
\newblock {\em Physical Review D}, 80(10):101701, 2009.

\bibitem{litim2001optimized}
Daniel~F Litim.
\newblock {O}ptimized renormalization group flows.
\newblock {\em Physical Review D}, 64(10):105007, 2001.

\end{thebibliography}

\end{document}